\documentclass[12pt]{article}
\usepackage{epsf}
\usepackage{amsmath}
\usepackage{graphics}

\begin{document}

\newcommand{\gtrsim}{ \mathop{}_{\textstyle \sim}^{\textstyle >} }
\newcommand{\lesssim}{ \mathop{}_{\textstyle \sim}^{\textstyle <} }

\newcommand{\sheptitle}
{Bigger Rip with No Dark Energy}
\newcommand{\shepauthor}
{Paul H. Frampton and Tomo Takahashi}
\newcommand{\shepaddress}
{Department of Physics and Astronomy,\\
University of North Carolina, Chapel Hill, NC 27599-3255.}
\newcommand{\shepabstract}
{}

\begin{titlepage}
\begin{flushright}
astro-ph/0405333 (updated)

\today
\end{flushright}
\vspace{.1in}
\begin{center}
{\large{\bf \sheptitle}}
\bigskip \medskip \\ \shepauthor \\ \mbox{} \\ {\it \shepaddress} \\
\vspace{.5in}

\bigskip \end{center} \setcounter{page}{0}
\shepabstract
\begin{abstract}
By studying a modified Friedmann equation which arises in
an extension of general relativity which accommodates a time-dependent
fundamental length $L(t)$, we consider cosmological models where
the scale factor diverges with an essential singularity at a finite
future time. Such models have no dark energy in the conventional sense
of energy possessing a truly simple pressure-energy relationship. Data on
supernovae restrict the time from the present until
the Rip to be generically longer than the current age of
the Universe.
\end{abstract}

\vspace*{5cm}
\begin{flushleft}
\hspace*{0.9cm}

\begin{tabular}{l} \\
\hline
{\small Emails:
frampton@physics.unc.edu,
takahash@physics.unc.edu}

\end{tabular}
\end{flushleft}

\end{titlepage}

\newpage

\section{Introduction}

The cosmic concordance of data from three disparate sources:
Cosmic Microwave Background (CMB), Large Scale Structure (LSS)
and Type Ia Supernovae (SNeIa) suggests that the
present values of the dark energy and matter components,
in terms of the critical density, are approximately
$\Omega_{\rm X} \simeq 0.7$ and $\Omega_{\rm m} \simeq 0.3$.

The equation of state of the dark energy $w = p/\rho$
suggests the possibility that $w < -1$, first studied
by Caldwell\cite{caldwell} and subsequently in a number 
of papers\cite{phantom,FT}.

The conclusion about the make-up of our Universe depends on assuming
that general relativity(GR) is applicable at the largest cosmological
scales. Although there is good evidence for GR at Solar-System
scales\cite{will} there is no independent evidence for GR at scales
comparable to the radius of the visible Universe. 
This involves an extrapolation in scale comparable
to that from the weak to the grand unified scale
in particle physics (some 13 to 25 orders of magnitude)
and could be called the cosmological desert hypothesis.
The expansion rate
of the Universe, including the present accelerating rate of cosmic
expansion can be parameterized in the right-hand side of the Friedmann
equation by including a dark energy density term with some assumed
time dependence on the scale parameter $a(t)$: $\rho_{DE} \sim
a^{\beta}$ with $\beta = - 3(1 + w)$. This is a rather restricted
function if we assume that the equation of state $w$ is
time-independent.  But as soon as we admit that it may depend on time
$w(t)$ then the function on the right hand side of the Friedmann
equation becomes completely arbitrary, just as does its designation as
a dark energy density. We restrict the designation ``dark energy"
for the normal case where
general relativity is assumed at all length scales
and the equation of state $w = p/\rho$ is a constant or
at most a very simple function of time.

Present data are fully consistent with constant $w=-1$ corresponding
to a cosmological constant which we may accommodate on the left hand
side of the Friedmann equation describing the expansion rate of the
Universe. But cases with $w \neq -1$, including $w < -1$, are still
permitted by observations.  In this case, there is a choice between
cooking up a ``dark energy'' density with a particular time dependence
$\rho_{DE} \sim a^{\beta(t)}$ on the right-hand side of the Friedmann
equation or changing the left-hand-side by changing the relationship
between the geometry and the matter density, {\it i.e.}  by changing
GR. 

The case constant $w < -1$ has the interesting outcome
for the future of the Universe that it will end in a finite
time at a ``Big Rip'' before which all structure disintegrates\cite{CKW}.

In the present article, we shall study an amalgam of the modification
of GR due to Dvali, Gabadadze and Porrati\cite{DGP} (DGP) and the idea
of a Big Rip, in fact here a Bigger Rip. The aim is mere;y to study
the range of possibilities in modified Friedmann
equations but the results are sufficiently interesting to examine and
such modifications may be constrained by observational data.

\section{Set up}

The DGP gravity\cite{DGP} arises from considering the four-dimensional
gravity which arises from five-dimensional general relativity confined
to a brane with three space dimensions. The underlying action is:

\begin{equation}
S = M(t)^3 \int d^5X \sqrt{G} {\cal R}^{(5)} +
M_{Planck}^2 \int d^4x \sqrt{g} R
\label{action}
\end{equation}
where ${\cal R}^{(5)}$ and $R$ are the scalar curvature in 5- and 4-
dimensional spacetime respectively, and $G$ and $g$ are the
determinant of the 5- and 4- dimensional spacetime metric.

This leads to an interesting modification of GR which embodies a
time-dependent length scale $L(t) = M_{Planck}^2/M(t)^3$.  For
cosmology it is natural to identify, within a coefficient of order
one, $L(t)$ with the Hubble length $L(t) = H(t)^{-1}$ at any
cosmological time $t$. Actually we are slightly generalizing the
original DGP approach to include time dependence of the length scale $L(t)$.

Taking the four dimensional coordinates to be labeled by
$i,k = 0, 1, 2, 3$ leads to the following modification
of Einstein's equation, as a result of varying the action Eq.(\ref{action}):

\begin{equation}
\left( R^{ik} - \frac{1}{2} R g^{ik} \right)
+ \frac{2 \sqrt{G}}{L(t) \sqrt{g}}
\left[ 
\left( {\cal R}^{(5)ik} - \frac{1}{2} G^{ik} {\cal R}^{(5)} \right) 
\right]
= 0
\label{Einstein}
\end{equation}
where the specialized notation [()] in Eq.(\ref{Einstein})
means the following:
\begin{equation}
\int dx [( f(x) )] \equiv f^{'}(0) \delta(x).
\label{distribution}
\end{equation}
and in Eq.(\ref{distribution}) we here identify $x$ with the additional space
dimension.

It is interesting to generalize the Schwarzschild solution to this
modification of GR\cite{gruzinov}. One finds that the modification of
the Newton potential at short distances is given by:

\begin{eqnarray}
V(r) & = & - \frac{Gm}{r} - \frac{4 \sqrt{Gm} \sqrt{r}}{L(t)} \notag \\ 
& = & - \frac{r_g}{2 r} - \frac{2 \sqrt{2} \sqrt{r_g r}}{L(t)}
\label{potential}
\end{eqnarray}
where $r_g = 2Gm$ is the Schwarzschild radius.

The fractional change in the Newtonian gravitational potential at
cosmological time $t$ at orbital distance $r$ from an object with
Schwarzschild radius $r_g$ is therefore

\begin{equation}
\left| \frac{\Delta V}{V} \right| = \sqrt{\frac{8 r^3}{L(t)^2 r_g}}
\label{DeltaV}
\end{equation} 

In the Bigger Rip scenario we will describe the characteristic length
$L(t)$ will decrease with time according to
\begin{equation}
L(t) = L(t_0) T(t)^p
\label{L}
\end{equation}
where the power satisfies $p \ge 1$ ($p < 1$ implies that $L(t)$
would {\it increase}) and where
\begin{equation}
T(t) = \frac{(t_{rip} - t)}{(t_{rip} - t_0)}
\label{T}
\end{equation}
in which $t_{rip}$ is the time of the Rip.  A bound system will become
unbound at a time $t_U$ when the correction to the Newtonian potential
becomes large. We make adopt the value of $t_U$ defined from
Eq.(\ref{DeltaV}) by
\begin{equation}
\sqrt{\frac{8 r^3}{L(t_U)^2 r_g}} = 1
\label{tU}
\end{equation}
We can rewrite Eq.(\ref{tU}) as:
\begin{equation}
(t_{rip}-t_U) = \frac{1}{\gamma} \left(
\frac{8 l_0^3}{L_0^2 r_g} \right)^{\frac{1}{2p}}
\label{tU2}
\end{equation}
where $\gamma = (t_{rip} - t_0)^{-1}$.

We shall define another later time $t_{caus}$ as the time after which
the two objects of a bound system become causally disconnected from
$t_{caus}$ until $t_{rip}$. This is defined by the equation:
\begin{equation}
(t_{rip} - t_{caus}) = \frac{l_0}{c} \left(
\frac{a(t_{caus})}{a(t_U)} \right)
\label{tcaus}
\end{equation}

\bigskip
\bigskip

As an example taking $p=1$ with the values $L_0 = H_0^{-1} =
(14Gy)^{-1} = 1.3\times 10^{28}cm$ and $\gamma = (20Gy)^{-1}$ we
arrive at the entries in the following Table:

\bigskip
\bigskip

\begin{tabular}{||c||c|c|c|c||}
\hline
Bound system & $l_0(cm) $ & $r_g$(cm) & 
$(t_{rip} - t_U)$ & $(t_{rip} - t_{caus})$  \\
\hline\hline
Typical galaxy & $5\times 10^{22}$ & $3\times10^{16}$ & 100My & 4My \\
\hline
Sun-Earth & $1.5 \times 10^{13}$ & $2.95 \times 10^5$ & 2mos & 31hr \\
\hline
Earth-Moon & $3.5\times 10^{10}$ & $0.866$ & 2weeks & 1hr \\
\hline
\hline
\end{tabular}

\bigskip
\bigskip
\noindent Note that the values we find for $(t_{rip} - t_U)$ are
consistent with those found in \cite{CKW}. The corresponding dark
energy would have equation of state $w = -1 - \frac{2}{3} \gamma L_0 =
- 1.466$ which, like that of \cite{CKW}, is now outside of the range
allowed by observations\cite{Riess} if we assume a constant equation
of state although it is allowed in the present model with its
time-dependence. As another example, with more normal present $w$, we
can increase the time to the Rip to $\gamma = (50Gy)^{-1}$ in which
case $w(t_0) = -1.19$ and the Table is modified to:

\bigskip
\bigskip

\begin{tabular}{||c||c|c|c|c||}
\hline
Bound system & $l_0(cm) $ & $r_g$(cm) & 
$(t_{rip} - t_U)$ & $(t_{rip} - t_{caus})$  \\
\hline\hline
Typical galaxy & $5\times 10^{22}$ & $3\times10^{16}$ & 250My & 7My \\ 
\hline
Sun-Earth & $1.5 \times 10^{13}$ & $2.95 \times 10^5$ & 5mos & 2days \\
\hline
Earth-Moon & $3.5\times 10^{10}$ & $0.866$ & 1mo & 2hrs \\
\hline
\hline
\end{tabular}

\bigskip
\bigskip

\noindent so with the more lengthy wait until the Big Rip the
disintegration of structure and causal disconnection occur
correspondingly earlier before the eventual Rip.

\bigskip
\bigskip

\section{The Bigger Rip}

\bigskip
\bigskip

The modified Friedmann equation for DGP gravity is
\begin{equation}
H^2 - \frac{H}{L(t)} = 0
\label{DGPFriedmann}
\end{equation}
so that we arrive at:
\begin{equation}
\frac{\dot{a}}{a} = H(t) = H(t_0)\frac{1}{T^p} 
\label{DGPFriedmann2}
\end{equation}
In Eqs.(\ref{DGPFriedmann},\ref{DGPFriedmann2}) we
can neglect, for the future evolution,
the term $(\rho_M + \rho_{\gamma})/(3 M_{Planck}^2)$
on the right-hand-side of
the modified Friedmann equation.
Defining $\gamma = -dT/dt = (t_{rip}-t_0)^{-1}$ gives:
\begin{equation}
{\rm ln} a(t) = - \int_{1}^{T(t)} \frac{dT}{\gamma L(t_0) T^p} 
\end{equation}
and hence, for $p=1$, which is similar to dark energy with a constant
$w < -1$ equation of state:
\begin{equation}
a(t) = T^{-\frac{1}{\gamma L(t_0)}}
\end{equation}
while for the Bigger Rip case $p>1$ one finds
\begin{equation}
a(t) = a(t_0) \exp \left[ \left( \frac{1}{T^{p-1}} - 1 \right) 
\frac{1}{(p-1)\gamma L(t_0)} \right]
\label{a(t)}
\end{equation}
Here we see that the scale factor diverges more singularly
in $T$ for $p>1$, hence the designation of {\it Bigger Rip}.
In particular we study the values $p = 2, 3, \cdots$ as alternative
to the ``dark energy'' case $p = 1$.

Inverting Eq.(\ref{a(t)}) gives:

\begin{equation}
T = [1 + (p-1) \gamma L(t_0) \ln a(t)]^{-\frac{1}{(p-1)}}
\label{T2}
\end{equation}
In this case there is strictly no dark energy, certainly not
with a constant equation of state, but we can mimic it with
a fictitious energy density $\rho_L$ by noticing that
$H^2 \sim T^{-2p}$
and writing 
\begin{equation}
\rho_L \sim [1 + (p-1) \gamma L(t_0) \ln a(t)]^{\frac{2p}{(p-1)}}
\label{rhoX}
\end{equation}
If we use Eqs.(\ref{a(t)}) and (\ref{rhoX}) in
conservation of energy
\begin{equation}
\frac{d}{dt} (\rho_L a^3) = -p \frac{d}{dt}(a^3) = -w_L(t) 
\rho_L \frac{d}{dt}(a^3)
\label{conservation}
\end{equation}
we find a time-dependent $w_L(t)$ for the ``fictitious'' dark energy
\begin{eqnarray}
w_L(t) & = & -1 + \frac{2}{3} \frac{dL(t)}{dt}  \\
& = & -1 -\frac{2}{3} \frac{p \gamma L(t_0)}{1+ (p-1)\gamma L(t_0) \ln a(t)}
\label{w(t)}
\end{eqnarray}
so the effective $w_L(t)$ has the limiting
values
$w_L(t_0) = -1 -\frac{2}{3}p(\gamma L(t_0))$ and $w_L(t_{rip}) = -1$.

We may check consistency with the space-space components of Einstein's
equations which are
\begin{equation}
\frac{\ddot{a}}{a} = -\frac{1}{2} \frac{\dot{a}^2}{a^2} - 4 \pi G p
\label{ii}
\end{equation}
which leads to
\begin{equation}
\dot{H} = - 4\pi G (\rho_L + p_L) = - 4 \pi \rho_L (1+w_L)
\label{ii2}
\end{equation}
so that with $H=L^{-1}$ and
$\rho_L = 3H/(8 \pi G L)$
we find a $w_L(t)$ consistent with Eq.(\ref{w(t)}).

\bigskip
\bigskip

Keeping the value $L_0 = H_0^{-1} = (14Gy)^{-1} = 1.3\times 10^{28}$
cm and putting $\gamma = (20Gy)^{-1}$ and $p=2$ we arrive at the
entries in the following Table:

\bigskip
\bigskip

\begin{tabular}{||c||c|c|c|c||}
\hline
Bound system & $l_0(cm) $ & $r_g$(cm) &
$(t_{rip} - t_U)$ & $(t_{caus} - t_{U})$  \\
\hline\hline
Typical galaxy & $5\times 10^{22}$ & $3\times10^{16}$ & $2.37Gy$ & 1.14Gy  \\
\hline
Sun-Earth & $1.5 \times 10^{13}$ & $2.95 \times 10^5$ & $9.6\times10^4y$ & $7y.$  \\
\hline
Earth-Moon & $3.5\times 10^{10}$ & $0.866$ & $2.5\times 10^4y$ & $6mos.$  \\
\hline
\hline
\end{tabular}

\bigskip
\bigskip

Note that for the $p=2$ case we have tabulated the difference
$(t_{caus}-t_{U})$ rather than $(t_{rip}-t_{caus})$ because in this
case the expansion is so rapid.

\bigskip
\bigskip

Next we turn to observational constraints on the parameters
$L_0$ and $\gamma$ for $p=2$.

\bigskip
\bigskip

\newpage

\bigskip
\bigskip

\section{Observational Constraints}

\bigskip
\bigskip

In the previous section, we discussed the future universe in the
model.  In this section, we discuss constraints on model parameters
from SNeIA data. To discuss the constraint, we have to include other
component such as cold dark matter (CDM) and baryon.  Including all
components, we can write the Friedmann equation as
\cite{Deffayet:2001pu}
\begin{equation}
H^2 + \frac{k}{a^2} = \left( 
\sqrt{ \frac{\rho_m}{3 M_{\rm Planck}^2} + \frac{1}{4 L^2}} 
+ \frac{1}{2L} \right)^2
\label{modFriedmann}
\end{equation}
If we define the density parameter $\Omega_m \equiv \rho_m /\rho_{\rm
crit} = \rho_{m0} (1+z)^3$, we can rewrite Eq.\ (\ref{modFriedmann})
as
\begin{equation}
H^2 = H_0^2 \left[ \Omega_k (1+z)^2 
+ \left( \sqrt{\Omega_L} 
+ \sqrt{\Omega_L + \Omega_m (1+z)^3} 
\right)^2 \right]
\end{equation}
where $\Omega_k$ and $\Omega_L$ are defined as 
\begin{equation}
\Omega_k \equiv \frac{-k}{H_0^2}, ~~~~~ 
\Omega_L \equiv \frac{1}{4 L^2 H_0^2}. 
\end{equation}
Thus at the present time, we have the relation among the density 
parameters,
\begin{equation}
\Omega_k + \left( \sqrt{\Omega_L} 
+ \sqrt{\Omega_L +  \Omega_m} 
\right)^2 =1.
\end{equation}

To obtain a constraint from the SNeIa observations, we assume that
red-shift dependence of $\Omega_L$ is as in Eq.\ (\ref{rhoX}) and that
$L$ is dependent on time as Eq.\ (\ref{L}); thus we can say that we
consider a new component $\rho_L$ which is defined as Eq.\
(\ref{rhoX}).

Now we discuss the constraint on this model from SNeIa data using
recent result \cite{Riess}.  In Fig.\ \ref{fig:fig1}, we show contours
of 95 and 99 \% C.L.  in $\Omega_L$-$\Omega_m$ plane for $\gamma L_0=
0$ (which is the constant $L$ case), 0.5 and 1 with $p=2$.  In the
figure, we also plot the line for the flat universe. Notice that the
line is different from the standard case because we have the modified
Friedmann equation in this model.  To obtain the constraint, we
marginalize the Hubble parameter dependence by minimizing $\chi^2$ for
the fit.

\bigskip
\bigskip

In Fig.\ \ref{fig:fig2}, we show the constraint on $\Omega_{\rm
m}$-$\gamma L_0$ plane assuming the flat universe ($\Omega_k=0$).  If
we take the value $\Omega_{\rm m}=0.3$, we can find an upper limit for
$\gamma L_0$ which is $\gamma L_0 \lesssim 0.7$.  This implies that
the time remaining from now until the Rip is constrained to be
generically at least somewhat longer than the current age.  If we make
$L_0$ larger than the length corresponding to the age of the Universe
then the upper bound on $\gamma$ diminishes and hence the time until
the Rip increases.

\bigskip

We note that because the effective equation of state $w(t)$ is varying
with time its present value $w(t_0)$ can be more negative than allowed
by constraints derived from assuming constant $w$.  Our constraint on
$\gamma L_0 < 0.7$ permits $w(t_0) = -1-\frac{2}{3} p \gamma L_0$ to
be as negative as $w(t_0) = - 1.9$ for $p=2$. Assuming constant $w$,
on the other hand, gives \cite{Riess} $w > -1.2$.

\newpage

\bigskip
\bigskip

\section{Discussion}

\bigskip
\bigskip

The present article is a natural sequel to our previous
paper \cite{FT} about dark energy in which it was pointed out that
no amount of observational data can, by itself, tell us
the fate of dark energy if we allow for an arbitrarily
varying equation of state. The three possibilities
listed were: there may firstly be a Big Rip,
or secondly dark energy may dominate but with an infinite lifetime
or thirdly the dark energy may eventually disappear leaving
a matter-dominated Universe. Given that observational data are insufficient,
only a successful and convincing theory may inform us confidently
of the future of the Universe.
 
\bigskip

The Big Rip was the most exotic of the fates and there seemed tied to
a phantom $w < -1$ dark energy. However, here we have studied
a Bigger Rip, in which the scale factor is even more
divergent at a future finite time than for the Big Rip, which
is achieved by modifying gravity and omitting dark energy.
In the model, as with the phantom case, structures become
unbound and subsequently their components become causally
disconnected before the Universe is torn apart in the Rip.

\bigskip

Work by other groups \cite{recent} has recently suggested 
that quite different values of cosmic parameters
can be acceptable if one relaxes the
most conventional and conservative assumptions of
general relativity at all length scales and a dark energy with
a constant equation of state.

\bigskip
\bigskip
\bigskip
\bigskip

\section*{Acknowledgments}

This work was supported in part by the US Department of Energy
under Grant No. DE-FG02-97ER-41036.


\bigskip
\bigskip

\newpage


\begin{thebibliography}{99}

\bibitem{caldwell}
R.R. Caldwell. 
Phys. Lett. {\bf B545,} 23 (2002).
{\tt astro-ph/9908168}.
\bibitem{phantom}
P.H. Frampton, Phys. Lett. {\bf B555}, 139-143 (2003).
{\tt astro-ph/0209037}\\
P.H. Frampton, Mod. Phys. Lett. {\bf A19,} 801 (2004).
{\tt hep-th/03020078}. \\
J.L. Crooks, J.O. Dunn, P.H. Frampton, H.R. Norton and T. Takahashi
Astropart. Phys. {\bf 20}, 361-367 (2003).
{\tt astro-ph/0305495.}\\
J.M. Cline, S.-Y. Jeon and G.D. Moore, Phys. Rev. {\bf D} (in press). {\tt hep-ph/0311312}.\\
M.P. Dabrowski, T. Stachowiak and M. Szydlowski, Phys. Rev. {\bf D68,} 103519 (203).\\
M. Kaplinghat and S. Bridle. {\tt astro-ph/0312430}.\\
S.M. Carroll, M. Hoffman and M. Trodden, Phys. Rev. {\bf D68,} 023509 (2003).
{\tt astro-ph/0301273}.
\bibitem{FT}
P.H. Frampton and T. Takahashi, Phys. Lett. {\bf B557}, 135-138 (2003). 
{\tt astro-ph/0211544}.
\bibitem{will}
C. M.  Will, {\it Was Einstein Right?: Putting General Relativity to the Test.}
Basic Books. Second Edition. (1993). 
\bibitem{CKW}
R.R. Caldwell, M. Kamionkowski and N.N. Weinberg, Phys. Rev. Lett. 
{\bf 91,} 071301 (2003). {\tt astro-ph/0302506}.
\bibitem{DGP}
G. Dvali, G. Gabadadze and M. Porrati,
Phys. Lett. {\bf B485,} 208 (2000). {\tt hep-th/0005016}.
\bibitem{gruzinov}
A. Gruzimov. {\tt astro-ph/0112246}.
\bibitem{Riess}
A.G. Riess {\it et al}, Ap. J. {\bf 607,} 665 (2004). {\tt astro-ph/0402512}.\\
A. Melchiorri, L. Mersini, C.J. Oldham and M. Trodden, Phys. Rev. {\bf D68,} 043509 (2003).
{\tt astro-ph/0211522}.
\bibitem{Deffayet:2001pu}
C.~Deffayet, G.~R.~Dvali and G.~Gabadadze,
Phys.\ Rev.\ D {\bf 65}, 044023 (2002).
{\tt astro-ph/0105068}.
\bibitem{recent}
S.K. Srivastava. {\tt astro-ph/0407048}.\\
E. Babichev, V. Dokuchaev and Yu. Eroshenko. {\tt astro-ph/0407190}.\\
S. Hannestad and E. Mortsell. {\tt astro-ph/0407259}.\\
B. A. Bassett, P.S. Corasaniti and M. Kunz. {\tt astro-ph/0407364}. \\
J.M. Virey {\it et al.} {\tt astro-ph/0407452}.
 

\newpage

\bigskip
\bigskip
\bigskip

\begin{figure}
    \centerline{\epsfxsize=1.5\textwidth\epsfbox{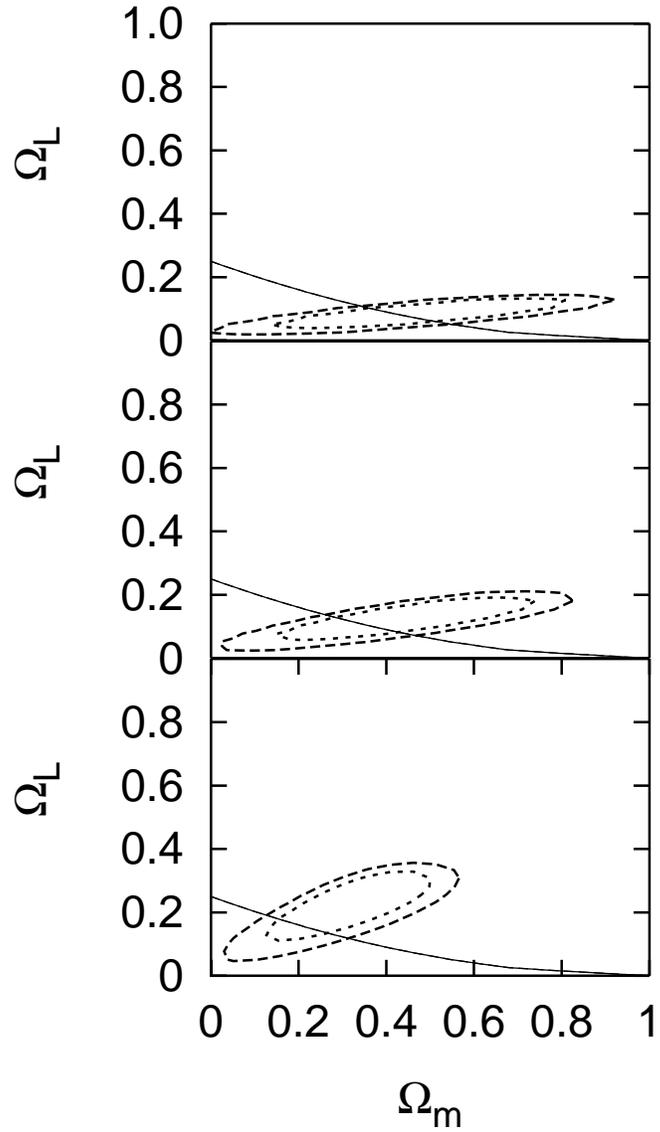}} 
\caption{Constraint from SNeIa observation in $\Omega_L$-$\Omega_{\rm
m}$ plane for $\gamma L_0=0$ (bottom), $\gamma L_0=0.5$ (middle) and
$\gamma L_0=2$ (top).  Contours are for 95 \% (dotted line) and 99 \%
(dashed line) C.L. constraints respectively. The solid line indicates
parameters which give a flat universe.}
   \label{fig:fig1}
\end{figure}

\newpage

\bigskip
\bigskip
\bigskip
\begin{figure}
    \centerline{\epsfxsize=1.\textwidth\epsfbox{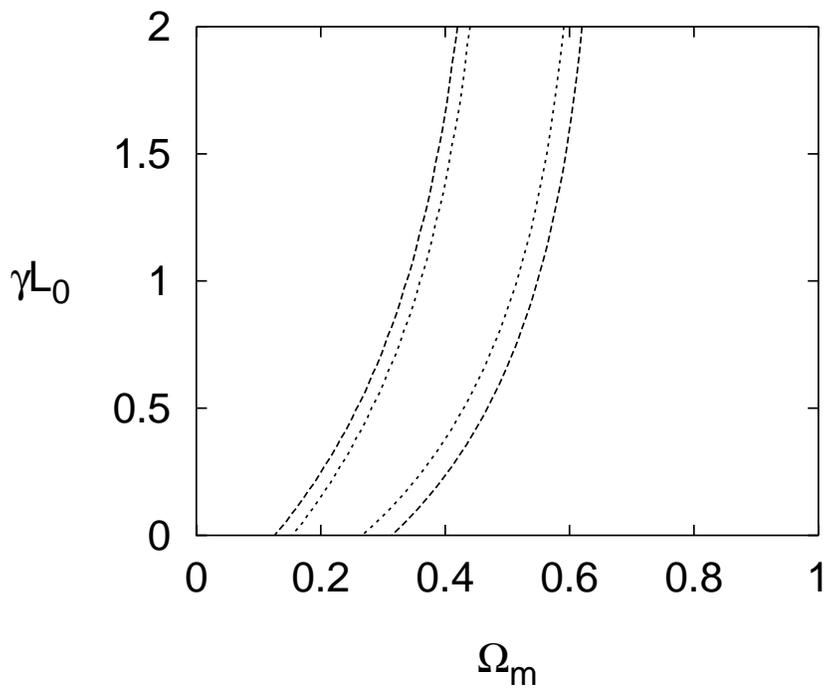}} 
\caption{Constraint from SNeIa observation in
$\Omega_{r_c}$-$\Omega_{\rm m}$ plane. Contours are for 95 \% (dotted
line) and 99 \% (dashed line) C.L. constraints respectively. In this
figure, we assume a flat universe.}
   \label{fig:fig2}
\end{figure}

\end{thebibliography}
\end{document}